\documentclass[aps,prd,showkeys,showpacs,superscriptaddress,floatfix]{revtex4}

\usepackage{graphics,epsfig}
\usepackage{epstopdf}
\usepackage{subfigure}
\usepackage{palatino}
\usepackage{changes}
\usepackage[colorlinks=true,linkcolor=blue,urlcolor=blue,citecolor=blue]{hyperref}
\usepackage[toc,page]{appendix}
\usepackage[normalem]{ulem}
\usepackage{adjustbox}
\usepackage{latexsym}
\usepackage{amsmath}
\usepackage{amssymb}
\usepackage{amsfonts}
\usepackage{times}
\usepackage{graphicx}
\usepackage{dcolumn}
\usepackage{amsmath}
\usepackage{array}
\usepackage{wrapfig}
\usepackage{multirow}
\usepackage{tabularx}
\begin{document}

\title{Effect of the magnetic charge on weak deflection angle and greybody bound of the black hole in Einstein-Gauss-Bonnet gravity}

\author{Wajiha Javed}
\email{wajiha.javed@ue.edu.pk; wajihajaved84@yahoo.com} 
\affiliation{Division of Science and Technology, University of Education, 54770 Township-Lahore, Pakistan}

\author{Muhammad Aqib}
\email{msf1900146@ue.edu.pk} 
\affiliation{Division of Science and Technology, University of Education, 54770 Township-Lahore, Pakistan}

\author{Ali {\"O}vg{\"u}n}
\email{ali.ovgun@emu.edu.tr}
\affiliation{Physics Department, Eastern Mediterranean University, Famagusta, 99628 North
Cyprus via Mersin 10, Turkey.}

\date{\today}

\begin{abstract}

The objective of this paper is to analyze the weak deflection angle of Einstein-Gauss-Bonnet gravity in the presence of plasma medium. To attain our results, we implement the Gibbons and Werner approach and use the Gauss-Bonnet theorem to Einstein gravity to acquire the resulting deflection angle of photon's ray in the weak field limit. Moreover, we illustrate the behavior of plasma medium and non-plasma mediums on the deflection of photon's ray in the framework of Einstein-Gauss-Bonnet gravity. Similarly, we observe the graphical influences of deflection angle on Einstein-Gauss-Bonnet gravity with the consideration of both plasma and non-plasma mediums. Later, we observe the rigorous bounds phenomenon of the greybody factor in contact with Einstein-Gauss-Bonnet gravity and calculate the outcomes, analyze graphically for specific values of parameters.
\end{abstract}



\maketitle

\section{Introduction}

In the theory of General Relativity (GR), black holes (BHs) are known as a great prediction of Einstein's theory of gravity and their geometrical characteristics are illustrated by simple mathematical equations. The most recent experimental observation is about the verification of the existence of gravitational waves \cite{1} and also the very first image of the supermassive BH shadow \cite{4} has allowed to recognize the behavior of the geometry and to analyze the various models in the context of strong field limit. Inspite of the fact that the Einstein theory of GR has the restriction of feasibility where its predictive power has been loosed in such regions. Therefore, its applicability and validity of higher order theories have been proposed to make the possible expressions of GR. Thus, the Lovelock theory has considered as an induction of Einstein theory \cite{6} as that is a higher order theory which consist of D-dimensions. Adding the Gauss-Bonnet (GB) Lagrangian to the Einstein-Hilbert action seems to be another way to compensate for higher order curvature terms.

In higher dimensions theory of gravity, lots of efforts have been made for the better understanding of the low-energy limit of String theory. An important higher dimensional generalization of Einstein gravity is the Einstein-Gauss-Bonnet gravity (EGBG) which is discovered by Lanczos in 1938 \cite{7}, and was rediscovered by LoveLock in 1971 \cite{6}. The study of EGBG becomes important since it provides a vast set up to explore a lot of conceptual issues related to the gravity.

The four dimensional EGBG theory, gets much attention in recent times \cite{8}-\cite{22}. The Glavan and Lin \cite{23} demonstrated the GB term in the EGBG theory as a topological invariant before regularization. 
After the LIGO detection of the gravitational waves, there is renewed interest in the topic of gravitational lensing \cite{1}.

Firstly gravitational lensing proposed by Soldner in 1801 in the context of Newtonian theory \cite{31}. As light emitted by distant galaxies, passes by massive objects in the universe, the gravitational pull from these objects can distort or bend the light's path. This is called \textit{gravitational lensing}. Three 
 types of gravitational lensing has been classified in the literature: (i) strong gravitational lensing (ii) weak gravitational lensing
(iii) micro-gravitational lensing \cite{30,32,33,35}. 

Gauss-Bonnet theorem is the most prominent technique that is used for calculating the weak deflection angle by means of optical geometry, which is engaged by Gibbons and Werner (GW) \cite{39,40}. The deflection angle calculated in the framework of GW is contemplated as a result of partially topological effect, the deflection angle is obtained by integrating the Gaussian optical curvature of the BH \cite{39}, which is given as follows
\begin{equation}
\breve{\alpha}=-\int \int_{D_\infty} \mathcal{K} dS\nonumber\\
\end{equation}
where $\breve{\alpha}$ denotes the deflection angle, $\mathcal{K}$ denotes the Gaussian optical curvature, $ds$ denotes the optical surface and $D_\infty$ symbolizes the infinite domain surrounded by the photon ray, apart from the lens. Thus, the GW methodology has discussed in a unique prospect for BHs and wormholes \cite{loboo}-\cite{67}.

According to Hawking's theory, BH is not perfectly "black" but instead actually emits particles. This radiation, could eventually loss the energy and mass of BHs to make them disappear. The Hawking radiations are changed due to the bending of continuum, while breeding to structurally infinity. The spill radiations are changed from at the geographically infinity radiations, the change can be analyzed by greybody factor \cite {74}. There are many many ways to find greybody factor, such as WKB approximation method \cite {75}-\cite{Ngampitipan:2013sf}. A blackbody that emits radiant energy and has same relative spectral energy distribution at the same temperature, the difference in the energy distributions in smaller amount, is called greybody factor. Mistry \textit{et al.} \cite{Mistry} have obtained the Hawking radiation power equations
in the background of asymptotically flat, AdS and dS BHs in $(d +1)$ dimensions by using the greybody factors for these BHs.

Main aim of the paper is to probe physical properties of the magnetic charge on the weak deflection angle and greybody bound of the black hole in Einstein-Gauss-Bonnet gravity. Astrophysically black holes can not be found alone but they are surrounded by some matter fields which can interact with them such as matter accretions, external magnetic fields. On the other hand, as NED is minimally coupled to Einstein-Gauss-Bonnet gravity, this magnetic charge term may have a significant contributions to deflection angle of the photon. Moreover, it can affect the photon rays in plasma medium. Next, it will
be also more interesting to see the contribution of the magnetic charge on the greybody factor of this black hole. To be specific, in the EGB gravity
black hole, we investigate the effects of magnetic charge parameter on the weak deflection angle which can be detected by astrophysical observations. Weak magnetic field is thought to influence the propagation of photons from a black hole to an
observer, and hence once can expect to find
correction to deflection angle in weak field limits which can be observed by experiments.

This work is characterized as follows; In section $\textbf{2}$, we inspect about EGBG. In section $\textbf{3}$, with the concern of the Gauss-Bonnet theorem, we probe the deflection angle in the context of non-plasma medium. In section $\textbf{4}$, we are concerning with the graphical analysis of deflection angle in non-plasma medium. In section $\textbf{5}$, we calculate the deflection angle for EGBG in plasma medium, and in section $\textbf{6}$ we analyze the graphical impact of deflection angle in the framework of plasma medium. In the section $\textbf{7}$, we enlarge the investigation and compute rigorous bound of greybody factor for EGBG and observe its graphical impact in last section.

\section{Weak Gravitational lensing and Einstein-Gauss-Bonnent Gravity}

The line-element of spherically symmetric $D$-Dimensional spacetime is defined as \cite{79};
\begin{equation}
 ds^2=-f(r)dt^2+ \frac{dr^2}{f(r)}+r^2d\Omega^2_{D-2}.\label{AH1}
\end{equation}
Here, $d\Omega_{D-2}^{2}$ denotes the line-element of the unit $(D-2)$-dimensional sphere
\begin{equation}
 f(r)=1-\frac{2GM}{r}+\frac{G q^2_{m}}{r^2}+\mathcal{O}(r^3)~~ r\gg1,
 ~~~d\Omega_{D-2} ^{2}=d\theta^2+\sin^2\theta d\varphi^2,
\end{equation}
here, the magnetic mass of BH is denoted by $M$, $r$ expresses the radial coordinate, $q^2_{m}$ represents magnetic charge and $G$ represents gravitational constant of BH. By placing the value of $f(r)$ in Eq.(\ref{AH1}), we attain the following equation
\begin{eqnarray}
ds^2&=&-\left(1-\frac{2GM}{r}+\frac{G q^2_{m}}{r^2}\right)dt^2+\frac{1}{\left(1-\frac{2GM}{r}+\frac{G q^2_{m}}{r^2}\right)} dr^2\nonumber\\&+&r^2d\theta^2+r^2\sin^2\theta d\varphi^2.\label{S2}
\end{eqnarray}
By making the assumption that the source and the viewer both are in the same equatorial plane along the pathway of null photons which is also on the equivalent plane keeping $(\theta=\frac{\pi}{2})$, for the sake of null geodesic we place $ds^{2}$=0 and we acquire the optical metric given as follows

\begin{eqnarray}
dt^2=\frac{1}{\left(1-\frac{2GM}{r}+\frac{G q^2_{m}}{r^2}\right)^{2}} dr^2+\frac{r^2}{\left(1-\frac{2GM}{r}+\frac{G q^2_{m}}{r^2}\right)} d\varphi^2.\label{S2}
\end{eqnarray}
 Afterwards, we make change in optical metric into a new frame of coordinate system $\tilde{r}$ composed as,
\begin{equation}
 dt^2=\bar{g}_{ab} dx^adx^b=d\tilde{r}^2 +f^{2}(\tilde{r})d\varphi^2,\label{AH2}
\end{equation}
 here
\begin{eqnarray}
 \tilde{r}&=&\frac{dr}{(1-\frac{2GM}{r}+\frac{G q^2_{m}}{r^2})},\nonumber\\
 f(\tilde{r})&=&\frac{r}{\sqrt{(1-\frac{2GM}{r}+\frac{G q^2_{m}}{r^2})}}.\label{AH3}
\end{eqnarray}
 It is notified that the previous system $(a,b)$ is transformed into new system $(r,\varphi)$ and determinant value is given as
 $det{\bar{g}_{ab}}=\frac{r^{2}}{f(\tilde{r})^{3}}$; make use of Eq.(\ref{AH2}), the value of remaining non-zero Christoffel symbols which are given as
 \begin{equation*}
 \Gamma^{\tilde{r}}_{\varphi \varphi}=\frac{r(rf'(\tilde{r})-2f(\tilde{r})}{2}~~~,~~~
 \Gamma^\varphi_{\tilde{r} \varphi}=\frac{-rf'(\tilde{r})+2f(\tilde{r})}{2rf(\tilde{r})}=\Gamma^\varphi_{\varphi \tilde{r}}~~~,~~~
 \Gamma^{\tilde{r}}_{\tilde{r} \tilde{r}}=-\frac{f'(\tilde{r})}{f(\tilde{r})}
 \end{equation*}
 and for optical curvature value of non-zero components of Riemann tensor is characterized as $R_{\tilde{r} \varphi \tilde{r} \varphi}$=$-k f^{2}(\tilde{r})$
 where $R_{\tilde{r} \varphi \tilde{r} \varphi}$=$g_{\tilde{r} \tilde{r}}
 R^{\tilde{r}}_{\varphi \tilde{r} \varphi}$. Hence, the Gaussian optical curvature $\mathcal{K}$ can be calculated by
\begin{equation}
 \mathcal {K}=\frac{R_{icciScalar}}{2}.
\end{equation}
 By taking into account the previous equation, the Gaussian optical curvature $\mathcal{K}$ in terms of radial coordinate $r$ of Schwarzschild metric can written as \cite{39}
\begin{equation}
\mathcal {K}=\frac{-1}{f(\tilde{r})}\left[\frac{dr}{d\tilde{r}}
\frac{d}{dr}(\frac{dr}{d\tilde{r}}) \frac{df}{dr}+ \frac{d^2f}
{dr^2}(\frac{dr}{d\tilde{r}})^2\right].\label{AH4}
\end{equation}
 Ultimately, the respective Gaussian optical curvature $\mathcal{K}$ of photon for Einstein-Gauss-Bonnet gravity can be calculated by using Eq.(\ref{AH3})
 into Eq.(\ref{AH4}), we obtain such a conclusion
 \begin{eqnarray}
\mathcal{K}&\thickapprox&-\frac{2GM}{r^3}+\frac{3Gq^2_{m}}{r^{4}}-\frac{6G^{2}Mq^2_{m}}{r^{5}}+\mathcal{O}(M^2,q^3_m,G^3).\label{AH6}
\end{eqnarray}

\section{Deflection Angle of Einstein-Gauss-Bonnet Gravity in Non-Plasma Medium}

 Now we find, the deflection angle for Einstein-Gauss-Bonnet gravity with the help of Gauss-Bonnet theorem in the framework of non-plasma medium.
 Hence, by utilizing the Gauss-Bonnet theorem to the domain $\mathcal{V}_{R}$, given as below \cite{39}
\begin{equation}
 \int\int_{\mathcal{V}_{R}}\mathcal{K}dS+\oint_{\partial\mathcal{V}_{R}}kdt
 +\sum_{z}\epsilon_{z}=2\pi\mathcal{F}(\mathcal{V}_{R}),
\end{equation}
 where $\mathcal{K}$ symbolizes the Gaussian optical curvature and $k$ symbolize geodesic curvature, characterized as
 $k=\bar{g}(\nabla_{\dot{\gamma}}\dot{\gamma},\ddot{\gamma})$ such as $\bar{g}
 (\dot{\gamma},\dot{\gamma})=1$, here unit acceleration vector designated by $\ddot{\gamma}$, the exterior angle of $z^{th}$ vertex symbolized as $\epsilon_{z}$. When $R$ goes to infinity, as a result the jump angle calculate by estimate to $\pi/2$ and we proceeds $\theta_{O}+\theta_{S}\rightarrow\pi$. Here, the value of Euler
 characteristic number is $(\mathcal{F}(\mathcal{V}_{R})=1)$ and $\mathcal{V}_{R}$ represents non-singular region. So, we conclude the following,
\begin{equation}
 \int\int_{\mathcal{V}_{R}}\mathcal{K}dS+\oint_{\partial
 \mathcal{V}_{R}}kdt+\epsilon_{z}=2\pi\mathcal{F}(\mathcal{V}_{R}),
\end{equation}
 Here, the total jump angle presented by $\epsilon_{z}=\pi$. When the value of $R\rightarrow\infty$, then the remainders of the part yield $k(E_{R})=\mid\nabla_{\dot{E}_{R}}\dot{E}_{R}\mid$. For the sake of geodesic curvature the value of radial component is given as,
\begin{equation}
 (\nabla_{\dot{E}_{R}}\dot{E}_{R})^{r}=\dot{E}^{\varphi}_{R}
 \partial_{\varphi}\dot{E}^{r}_{R}+\Gamma^{{\tilde{r}}}_{\varphi\varphi}(\dot{Z}^{\varphi}_{R})^{2}.\label{AH5}
\end{equation}
 By keeping $R$ is very high, then $E_{R}:=r(\varphi)=R=const$. Thus, the composition of the
  Eq.(\ref{AH5}) converts to $(\dot{E}^{\varphi}_{R})^{2}
 =\frac{1}{f^2(\tilde{r})}$. Memorizing $\Gamma^{\tilde{r}}_{\varphi \varphi}=\frac{r(rf'(\tilde{r})-2f(\tilde{r})}{2}$, we yield
\begin{equation}
 (\nabla_{\dot{E}^{r}_{R}}\dot{E}^{r}_{R})^{r}\rightarrow\frac{1}{R}.
\end{equation}
  So, $k(E_{R})\rightarrow \frac{1}{R}$. Using optical metric Eq.(\ref{AH2}), it can be written as $dt=Rd\varphi$. Hence;
\begin{equation}
 k(E_{R})dt=d\varphi.
\end{equation}
 All of the above results taking into count, we acquire
\begin{equation}
 \int\int_{\mathcal{V}_{R}}\mathcal{K}ds+\oint_{\partial \mathcal{V}_{R}} kdt
 =^{R \rightarrow\infty }\int\int_{M_{\infty}}\mathcal{K}dS+\int^{\pi+ \tilde{\delta}}_{0}d\varphi.\label{hamza2}
\end{equation}
 The photon ray at $0^{th}$ order in weak field deflection limit is determined as $r(t)=b/\sin\varphi$. Thus, utilizing
the above equations: then, the deflection angle determined as \cite{40};
\begin{equation}
 \tilde{\delta}=-\int^{\pi}_{0}\int^{\infty}_{b/\sin\varphi}\mathcal{K}\sqrt{det\bar{g}}d\tilde{r}d\varphi,\label{AH7}
\end{equation}
 where
\begin{equation}
 \sqrt{det\bar{g}}=r(1+\frac{3GM}{r}-\frac{3G q^2_{m}}{2r^2}).
\end{equation}
 By using the values of Gaussian curvature into Eq.(\ref{AH7}). Then deflection angle is calculated as:
\begin{equation}
\tilde{\delta} \thickapprox \frac{4GM}{b}-\frac{8G^{2} M q^{2}_{m}}{3b^{3}}-\frac{3G \pi q^{2}_{m}}{4b^{2}}+\mathcal{O}(M^2,q^3_m,G^3)\label{S1}
\end{equation}
The deflection angle $(\tilde{\delta})$ obtained in non-plasma medium depends on the mass $M$ of the BH, magnetic charge $q_{m}$ and impact parameter $b$. We  observe that in the attained deflection angle the first term is the well known result for the Schwarzschild BH and the other terms are due to the charge nature of the BH. It is also noted that the negative sign in the second term indicates that the weak bending angle of this magnetically charged BH is smaller than the Schwarzschild BH \cite{Qi}.

\section{Graphical Inspection for non-plasma medium}
 In this section, we observe the graphical influence of deflection angle. And also seek know about the vital importance of these graphs. Moreover, we investigate the impact of magnetic charge $(q_{m})$ and impact parameter $(b)$ on the deflection angle.

\subsection{Comparison between deflection angle $(\tilde{\delta})$ and impact parameter $(b)$}

\begin{center}
\epsfig{file=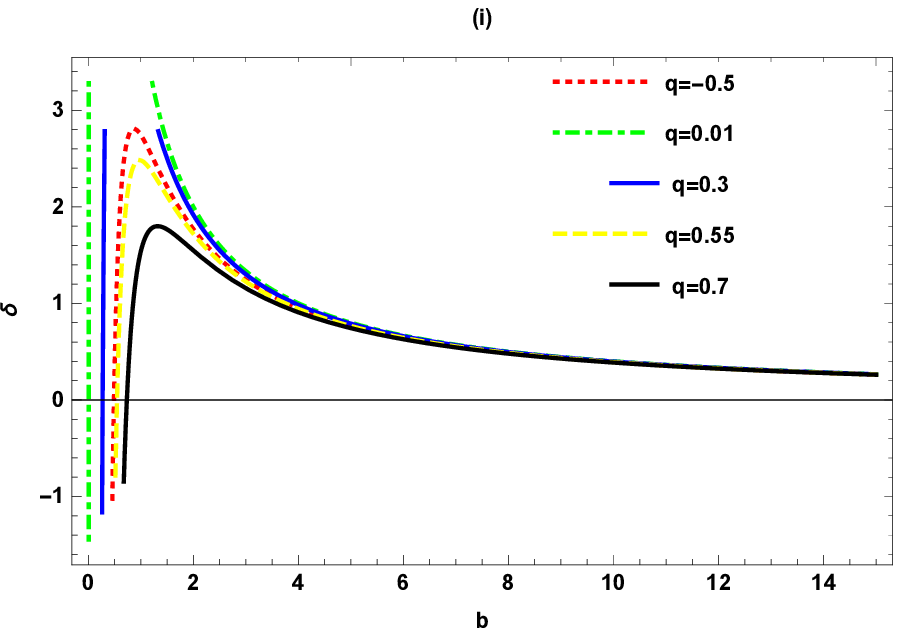,width=0.50 \linewidth}\epsfig{file=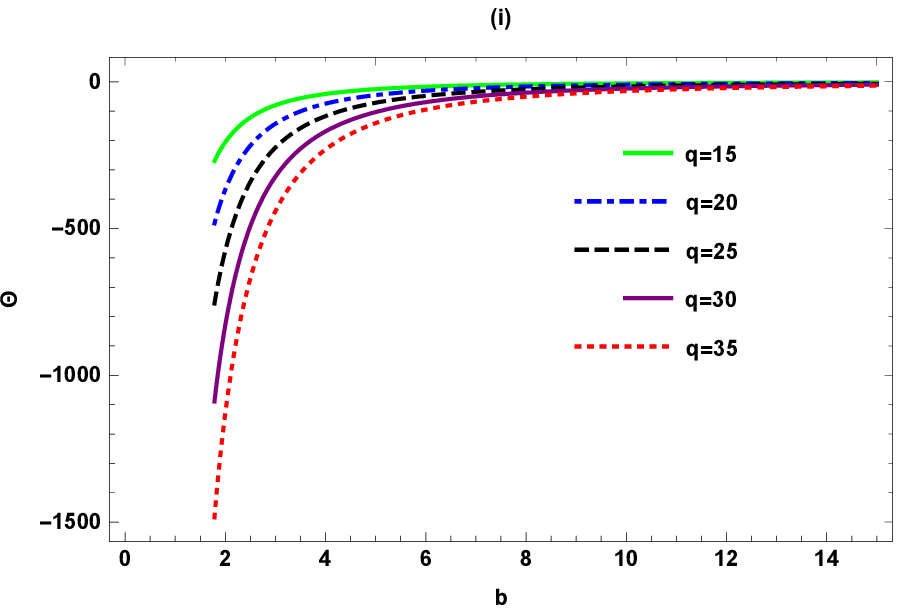,width=0.50 \linewidth}\\
{Figure 1: Correspondence between $\tilde{\delta}$ and $b$}.
\end{center}
\begin{itemize}
\item \textbf{Figure 1} Here, we can see the behavior of $\tilde{\delta}$ with respect to $b$ in order to varying the value of magnetic charge $q_{m}$ and remain both $G$ and $M$ fixed.
\begin{enumerate}
\item In left panel (i), we examined that the deflection angle $\tilde{\delta}$ gradually increase first for lower values of magnetic charge $q_{m}$ and at the end slope decrease.
\item In right panel (i), we observed that angle shows the decline behavior at higher values of magnetic charge $q_{m}$.
\end{enumerate}
\end{itemize}
\subsection{Comparison between deflection angle $(\tilde{\delta})$ and the magnetic charge $(q_{m})$}

\begin{center}
\epsfig{file=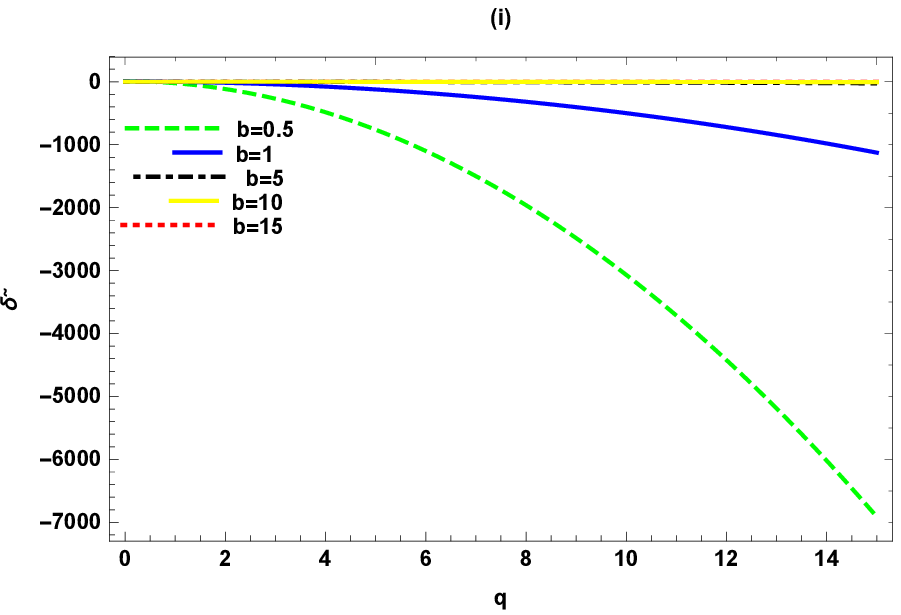,width=0.50\linewidth}\\
{Figure 2: Correspondence between $\tilde{\delta}$ and $q_{m}$}.
\end{center}
\begin{itemize}
\item \textbf{Figure 2} Now, we check the behavior of $\tilde{\delta}$ with respect to magnetic charge $q_{m}$ by changing $b$ and $G$,$M$ being fixed.
\begin{enumerate}
\item Figure (i), shows that the value of $\tilde{\delta}$ continually decrease by using the large values of $b$.
    \end{enumerate}
\end{itemize}

\section{Gravitational lensing for Einstein-Gauss-Bonnet Gravity in plasma medium}
 Now, this section is devoted to calculate the weak gravitational lensing for Einstein-Gauss-Bonnet gravity in the framework of plasma medium. For (EGB) gravity the value of refractive index $n(r)$ \cite{80}, given as;
\begin{equation}
 n^2\left(r,\omega(r)\right)=1-\frac{\omega_e^2(r)}{\omega_\infty^2(r)}.
 \end{equation}
 where the term $(\omega _{e})$ characterize for electron plasma frequency and the term $(\omega_\infty)$ express the photon frequency which is observed by a viewer at infinity, then the relevant optical metric presented as \cite{80};
\begin{equation}
 dt^2=g^{opt}_{ef}dx^pdx^f=n^2 \left[\frac{dr^2}{f^2(r)}+\frac{r^2d\varphi^2}{f(r)}\right],\label{hamza3}
\end{equation}
 with the value of determinant $g^{opt}_{ef}$,
\begin{equation}
\sqrt{g^{opt}}=r(1-\frac{\omega_e^2}{\omega_\infty^2})+GM(3
 -\frac{\omega_e^2}{\omega_\infty^2})-\frac{G q^2_{m}}{2r}(3
 -\frac{\omega_e^2}{\omega_\infty^2}).
\end{equation}

 Expression for Gaussian curvature in the form of curvature tensor written as;
\begin{equation}
    \mathcal{K}=\frac{R_{r\varphi r\varphi}(g^{opt})}{det(g^{opt})},\label{hamza1}
\end{equation}
 To applying Eq.(\ref{hamza1}), the following Gaussian curvature is calculated as,
\begin{eqnarray}
\mathcal{K}&\thickapprox&\frac{GM}{r^3}(-2-3\frac{\omega_e^2}{\omega_\infty^2}+4\frac{\omega_e^4}
{\omega_\infty^4})+\frac{G^{2}Mq^2_{m}}{r^5}(-6-26\frac{\omega_e^2}{\omega_\infty^2}
+28\frac{\omega_e^4}{\omega_\infty^4})\nonumber\\&+&\frac{Gq^2_{m}}{r^4}(3+5\frac{\omega_e^2}{\omega_\infty^2}-3\frac{\omega_e^4}{\omega_\infty^4})+\mathcal{O}(M^2,q^3_m,G^3)
\end{eqnarray}
 To do so, we make use of the (GBT) to find out deflection angle but also compare it with the non-plasma. Thus, for evaluating the deflection angle in the framework of weak field limit, as photon beams behaves like a straight line. Consequently, the limitation at $0^{th}$ order is $ r=\frac{b}{sin\varphi}$. So, the (GBT) is comprises into the following form to calculate the deflection angle $\tilde{\delta}$;
\begin{equation}
    \tilde{\delta}=-\lim_{R\rightarrow 0}\int_{0} ^{\pi} \int_\frac{b}{\sin\varphi} ^{\infty} \mathcal{K} dS
\end{equation}
 Using the above equation, so the value of deflection angle for Einstein-Gauss-Bonnet gravity in the framework of plasma medium is computed as;
\begin{eqnarray}
\tilde{\delta}&\thickapprox&\frac{4GM}{b}+\frac{GM}{b}(2\frac{\omega_e^2}{\omega_\infty^2}-14\frac{\omega_e^4}
{\omega_\infty^4})-\frac{8G^{2}Mq^2_{m}}{3b^3}+\frac{G^{2}Mq^2_{m}}{b^3}(2\frac{\omega_e^2}{\omega_\infty^2}-\frac{130\omega_e^4}
{9\omega_\infty^4})\nonumber\\&-&\frac{3G q^2_{m}\pi}{4b^2}+\frac{G q^2_{m}\pi}{b^2}(-\frac{\omega_e^2}{2\omega_\infty^2}+\frac{2\omega_e^4}
{\omega_\infty^4})+\mathcal{O}(M^2,q^3_m,G^3) ~~\label{S3}
\end{eqnarray}
The deflection angle $(\tilde{\delta})$ depends on the mass $M$ of the BH, magnetic charge $q_{m}$, impact parameter $b$ and on the plasma term. The bending angle obtained in the plasma medium increases with the parameter $\frac{\omega_e^2}{\omega_\infty^2}$, which shows that lower the photon frequency observed by a static spectator  at infinity, greater the deflection angle of it for the fixed electron plasma frequency. We also observe that when we take the magnetic charge $q_{m}=0$, the deflection angle obtained in the plasma medium reduces to the deflection angle of the Schwarzschild BH in plasma medium. We also investigate that the deflection angle obtained in the plasma medium reduces to the deflection angle that we have obtained in case of non-plasma, when we take $\frac{\omega_e^2}{\omega_\infty^2}=0$.

\section{Graphical Inspection for plasma medium}
   This section is presented the graphical impression of the deflection angle for Einstein-Gauss-Bonnet gravity. We prosecute the effect of numerous parameters on the deflection angle. Furthermore, we also elaborate the physical significance of these types of graph to inspect the actions of plasma medium. For our simplicity  we assign $G$=1,~$M=1$,   $\frac{\omega_e}{\omega_\infty}$=$10^{-1}$ and also vary the values of magnetic charge $q_{m}$ and impact parameters $b$ in order to obtain these graphs.

\subsection{Comparison between deflection angle $\tilde{\delta}$ and Impact parameter $b$}

\begin{center}
\epsfig{file=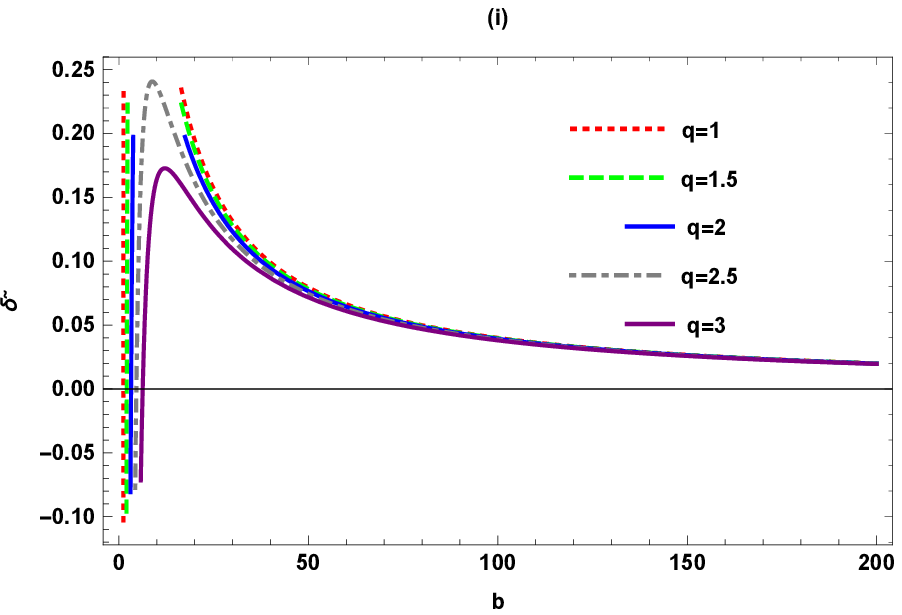,width=0.50\linewidth}\epsfig{file=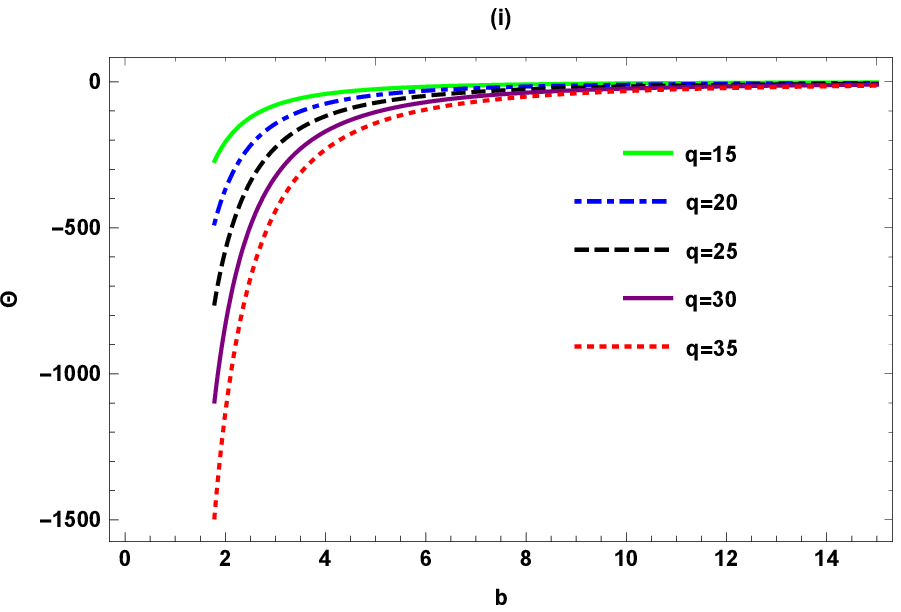,width=0.50\linewidth}\\
{Figure 3: Correspondence between $\tilde{\delta}$ and $b$}.
\end{center}

\begin{itemize}
\item \textbf{Figure 3} Represents the impact of $\tilde{\delta}$ with relates to impact parameter $b$ and vary the magnetic charge $q_{m}$ and taking the value of $M$, $G$ been constant.
\begin{enumerate}
\item In left fig(i), we can viewed that the value of deflection angle $\tilde{\delta}$ gradually increased first by putting different lower values of magnetic charge $q_{m}$ and then tends to move positive infinity.
\item In right fig(i), we can analyze that angle value decrease at the higher inputs of magnetic charge $q_{m}$.
\end{enumerate}
\end{itemize}

\subsection{Comparison between the deflection angle $\tilde{\delta}$ and magnetic charge $q_{m}$}

\begin{center}
\epsfig{file=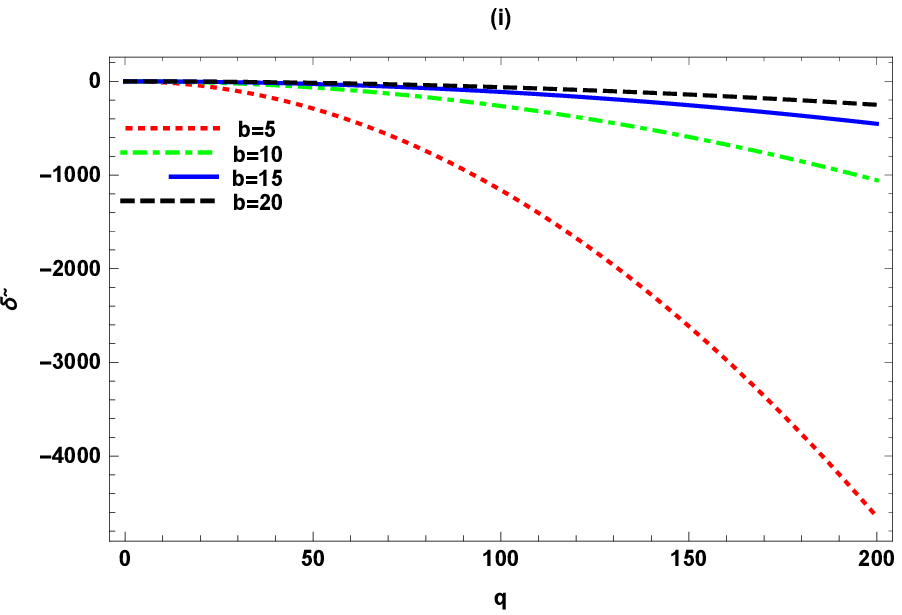,width=0.50\linewidth}\\
{Figure 4: Correspondence between $\tilde{\delta}$ and $q_{m}$}.
\end{center}
\begin{itemize}
\item \textbf{Figure 4} To show the correspondence of $\tilde{\delta}$ with magnetic charge $q_{m}$ and alternating the values of impact parameter and being fixed the values of $G$ and $M$ respectively.
\begin{enumerate}
\item In figure (i), we obtained that firstly deflection angle $\tilde{\delta}$ showed the straight line path and then steadily moves to negative infinity by changing values of $b$.
\end{enumerate}
\end{itemize}

\section{Greybody Factor For Einstein-Gauss-Bonnet-Gravity}
This section is based on the calculation of the rigrous bound of the greybody factor. The bound of greybody factor can be stated as \cite{83}
\begin{equation}
    T\geq sech^2\left(\frac{l}{2\omega}\int_{-\infty} ^{\infty}{\mathcal{V}(r)}dr_*\right)\label{T1}.
\end{equation}

The D-dimension spherically symmetric line element is given as \cite{79};
\begin{equation}
 ds^2=-f(r)dt^2+ \frac{dr^2}{f(r)}+r^2(d\theta^2+\sin^2\theta d\varphi^2),
\end{equation}
where,
\begin{equation}
f(r)=1-\frac{2GM}{r}+\frac{G q^2_{m}}{r^2}\nonumber\\
\end{equation}
The interior and exterior two event horizon value $r_\pm$ is given by
\begin{equation}
r_+=GM+\sqrt{G^{2}M^{2}-Gq^2_{m}},\nonumber\\
\end{equation}
\begin{equation}
r_-=GM-\sqrt{G^{2}M^{2}-Gq^2_{m}},
\end{equation}

The Schrodinger-like equation has been illustrated as;

\begin{eqnarray}
\left(\frac{d^{2}}{dr^{2}_{*}}+\omega^{2}-\tilde{V(r)}\right)\psi=0
\end{eqnarray}
here, $r^{2}_{*}$ denote the "tortoise coordinate".\\
\begin{equation}
dr^{2}_{*}=\frac{1}{f(r)}dr
\end{equation}
and

\begin{eqnarray}
\tilde{V(r)}=\frac{(d-2)(d-4)}{4}\frac{f^{2}(r)}{r^{2}}+\frac{(d-2)}{2}\frac{f(r)f'(r)}{r}+l(l+d-3)\frac{f(r)}{r^{2}}
\end{eqnarray}

In order to calculate the lower bound value on transmission probability taking the value $h=\omega$ is;\\
 \begin{eqnarray}
T&\geq&\frac{1}{\cosh^{2}}\left(\frac{1}{2\omega}\int^{\infty}_{-\infty}\tilde{V(r)dr^{2}_{*}} \right)\nonumber\\
&=&\frac{1}{\cosh^{2}}\left[\frac{1}{2\omega}\int^{\infty}_{r_+}\left(\frac{(d-2)(d-4)}{4}\frac{f(r)}{r^{2}}
+\frac{(d-2)}{2}\frac{f'(r)}{r}+\frac{l(l+d-3)}{r^{2}}\right)\right]\nonumber\\
&=&\frac{1}{\cosh^{2}}\left[\frac{1}{2\omega}\left(\frac{(d-2)}{2}\left(\frac{GM}{r^{2}_{+}}-\frac{2Gq^{2}_{m}}{3r^{3}_{+}}\right)+\frac{l(l+1)}{r_+}\right)\right]\nonumber
\end{eqnarray}
If we take $d=4$ and put $r_+$ value, then this bound value is reduced into the following
\begin{eqnarray}
T&\geq&\frac{1}{\cosh^{2}}\left[\frac{1}{2\omega}\left(\frac{GM}{(GM+\sqrt{G^{2}M^{2}-Gq^{2}_{m}})^{2}}-\frac{2Gq^{2}_{m}}{3(GM+\sqrt{G^{2}M^{2}-Gq^{2}_{m}})^{3}}
\right.\right.\nonumber\\&+&\left.\left.\frac{l(l+1)}{(GM+\sqrt{G^{2}M^{2}-Gq^{2}_{m}})}\right)\right]\nonumber
\end{eqnarray}
So, we have calculated the lower bound for EGBG. The obtained greybody bound of the BH depends on the mass $M$ of the BH and magnetic charge $q_{m}$.
We also observe that the greybody bound of the BH is similar to the bound of the Reissner-Nordstrom BH \cite{Ngampitipan:2013sf} when the electric charge is replaced by the magnetic charge.
If the BHs have no magnetic or electric charges, then the bound which is given above is reduced into
\begin{eqnarray}
T\geq\frac{1}{\cosh^{2}}\left[\frac{2\ell(\ell+1)}{8\omega GM}\right]
\end{eqnarray}
which is the same as the bound for the $4D$ Schwarzschild (BHs) emitting spinless particles \cite{83}.

\section{Graphical Behavior of Greybody Factor}
This section shows the graphical influences of greybody factor lower bound for (EGBG) and its potential while taking $(G=M=1)$ and different values of magnetic charge $q_m$ with angular momenta $l=0,1,2$ respectively.\\
\begin{center}
\epsfig{file=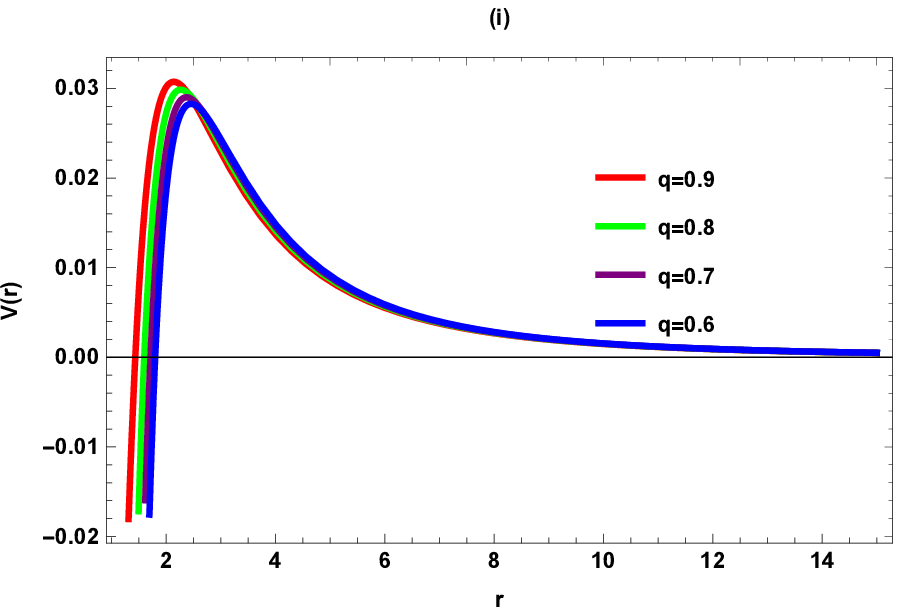,width=0.50\linewidth}~~\epsfig{file=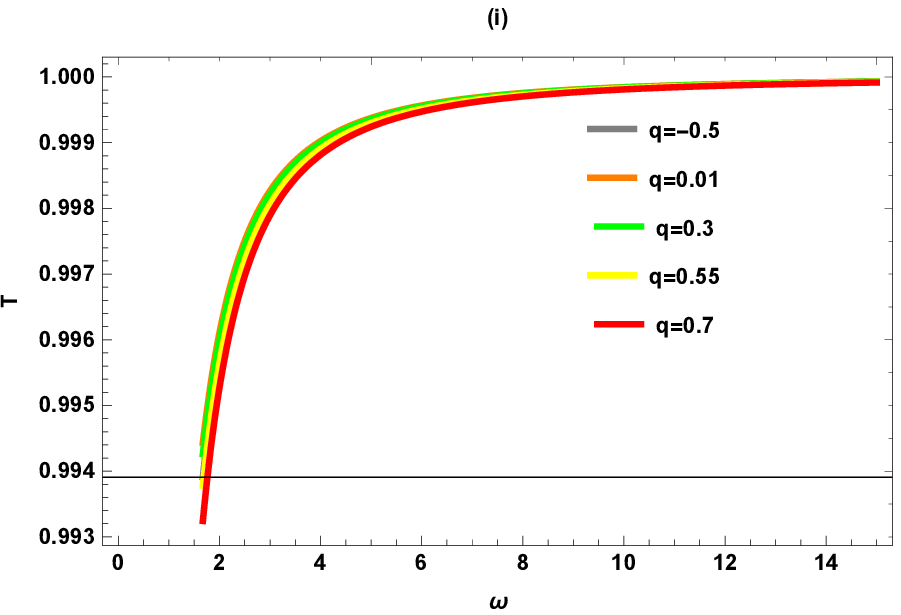,width=0.50\linewidth}\\
\epsfig{file=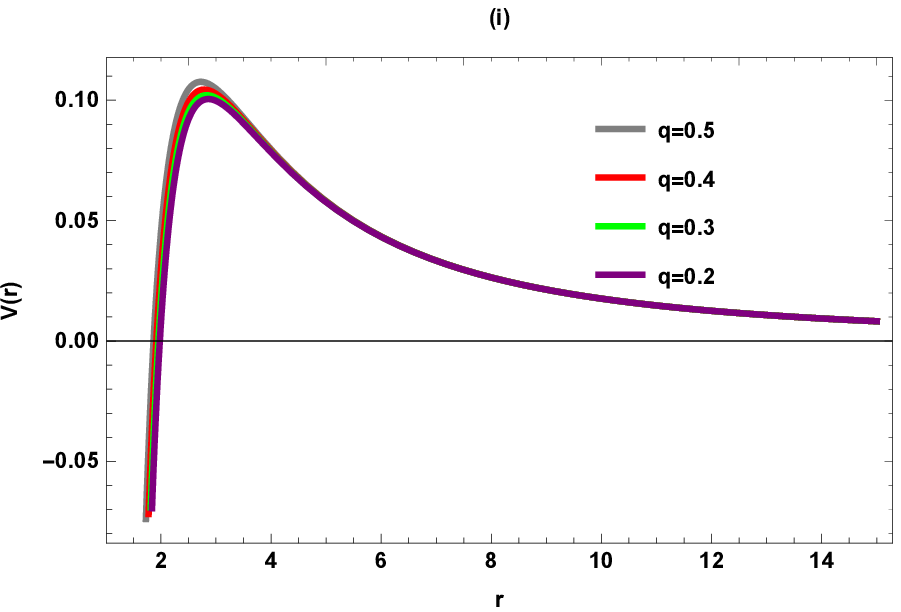,width=0.50\linewidth}~~\epsfig{file=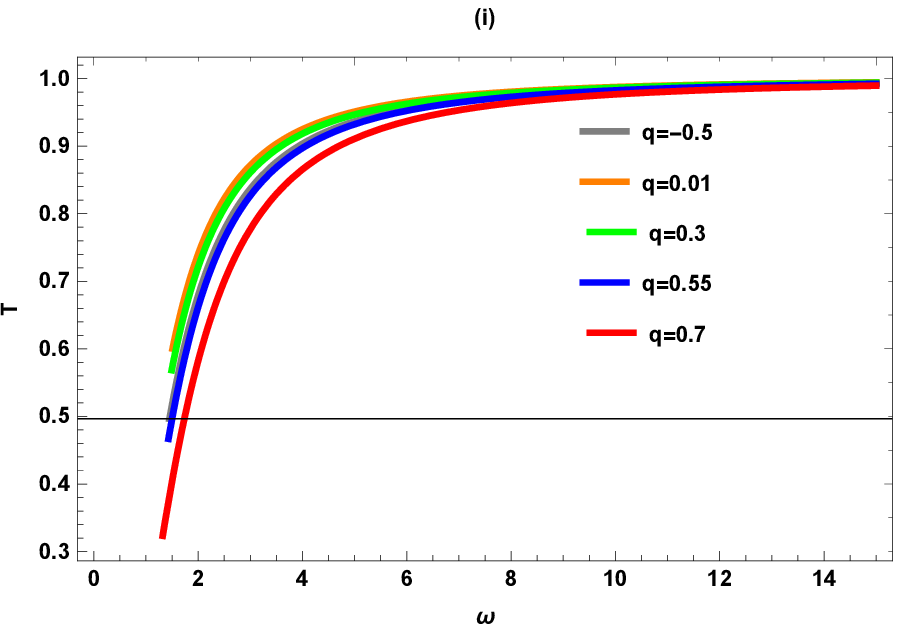,width=0.50\linewidth}\\
\epsfig{file=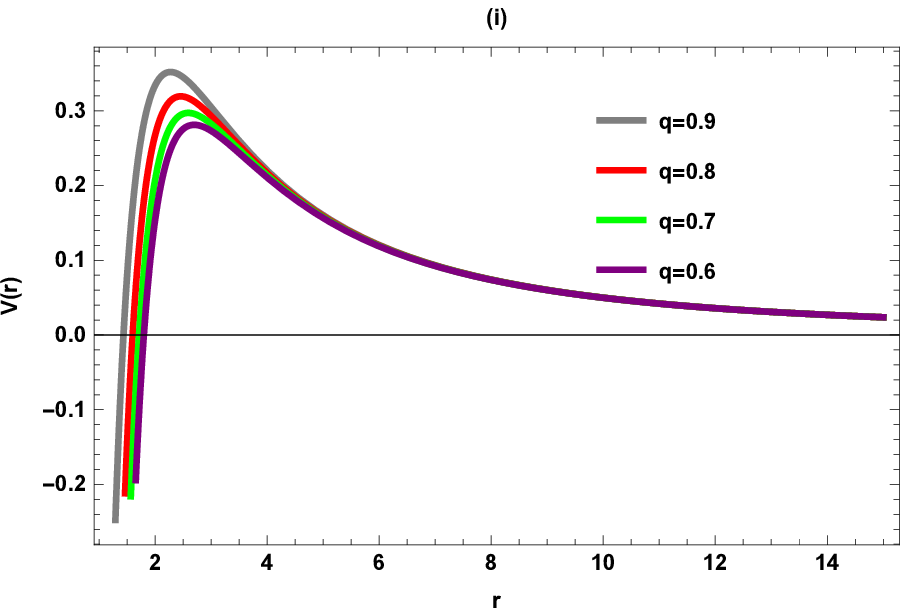,width=0.50\linewidth}~~\epsfig{file=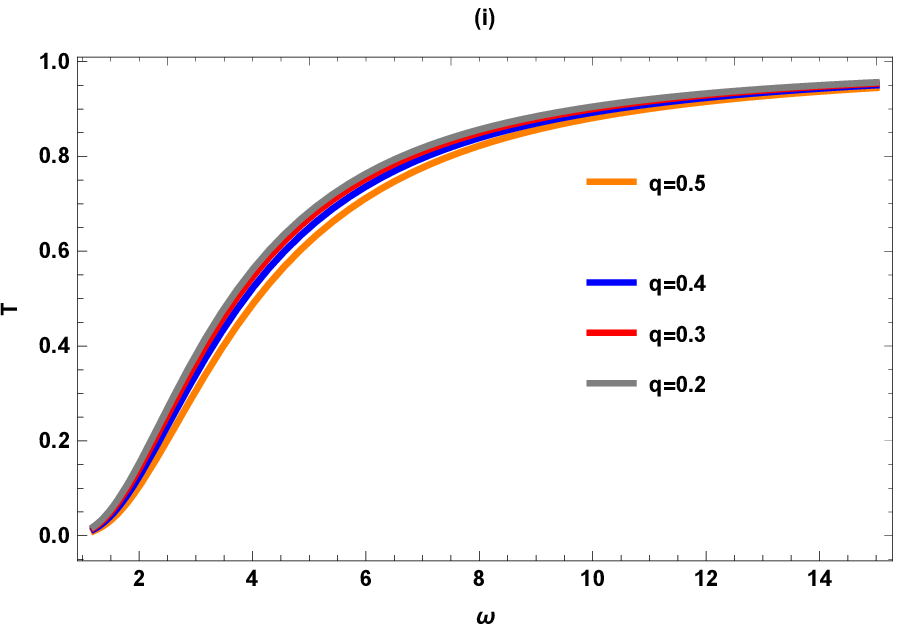,width=0.50\linewidth}\\
{Figure 5: The left figure corresponds the potential and the right figure corresponds the relevant greybody factor lower bound for (EGBG).}
\end{center}

Now, we can examine how $T$ behavior depends upon the potential shape. This analysis can be achieved by altering the magnetic charge parameter, $q_m$ as well as the angular momenta $\ell$. By fixing both $M=1$ and $G=1$, the potential takes the higher amplitude when $q_m$ changes as shown by the left plot in Figure.5. The greybody factor decrease for a certain value of $\omega$ whereas it is more difficult for the wave to be transmitted through higher potential value as shown by the right plot of Figure.\textbf{5}.

\section{Conclusion}

 This recent work is about to investigate deflection angle for Einstein-Gauss-Bonnet gravity in both cases for plasma and non-plasma mediums. To do so, we evaluate the weak lensing by applying (GBT) and derive deflection angle of photons ray for (EGB) gravity. The resulted deflection angle is;
\begin{equation}
   \tilde{\delta} \thickapprox \frac{4GM}{b}-\frac{8G^{2} M q^{2}_{m}}{3b^{3}}-\frac{3G \pi q^{2}_{m}}{4b^{2}}+\mathcal{O}(M^2,q^3_m,G^3)\label{S1}\nonumber\\
\end{equation}
  We investigate whether the resulted deflection angle can be reduced by reducing some parameters which turned into the deflection angle of Schwarzschild (BH) upto the first order value. Also, we review the graphical impression of alternate parameters on the deflection angle for Einstein-Gauss-Bonnet gravity. Further, the value of deflection angle is examined in the occupancy of plasma medium which is given by;
\begin{eqnarray}
\tilde{\delta}&\thickapprox&\frac{4GM}{b}+\frac{GM}{b}(2\frac{\omega_e^2}{\omega_\infty^2}-14\frac{\omega_e^4}
{\omega_\infty^4})-\frac{8G^{2}Mq^2_{m}}{3b^3}+\frac{G^{2}Mq^2_{m}}{b^3}(2\frac{\omega_e^2}{\omega_\infty^2}-\frac{130\omega_e^4}
{9\omega_\infty^4})\nonumber\\&-&\frac{3G q^2_{m}\pi}{4b^2}+\frac{G q^2_{m}\pi}{b^2}(-\frac{\omega_e^2}{2\omega_\infty^2}+\frac{2\omega_e^4}
{\omega_\infty^4})+\mathcal{O}(M^2,q^3_m,G^3)  ~~\label{S3}
\end{eqnarray}
 When the value of plasma effect $\frac{\omega_e}{\omega_\infty}$ nearest to zero, then the effect of plasma removed.
 Moreover, the graphical behavior of the deflection angle for Einstein-Gauss-Bonnet gravity in plasma comparison with some parameters. We examined the greybody factor for (EGBG) with the help of rigorous bound. Firstly, we compute the horizon value and then using Schrodinger-like equation which is obtained from the radial part, of the solution. Consequently, we observe the behavior of potential in order to investigate the greybody factor. It is conclude that slope height of the potential decrease when lower the magnetic charge values. Further, the rigorous bound upon greybody factors have been computed. It is discovered that we analyze qualitatively greybody factor bound by utilizing the following potential form; At high potential value, the wave face difficulty to be transmitted which cause the lower the greybody factor bound. The conclusion achieved from the inspection of deflection which is given in the recent work are proposed as follows;

\textit{\textbf{Deflection angle concern with impact parameter $b$:}}

 \begin{enumerate}
\item In our conclusion, we seen that firstly the value of deflection angle increasing and then after turned to decreasing for the smaller values of magnetic charge $q_{m}$.

\end{enumerate}
 \textit{\textbf{Deflection angle concern with magnetic charge $q_{m}$:}}

 \begin{enumerate}
\item We reviewed that the deflection angle firstly follows the straight path and then after tends to negative infinity for the values of impact parameter in range between $1<b<20$.
\end{enumerate}

According to our findings, the deflection angle of the magnetically charged black hole in EGB gravity is only slightly smaller than the uncharged black hole itself. However, this behaviour and further findings of the magnetic charges  can be best explained with an interesting assumption from observations in future. As a final remark, we can say that gravitational
lensing can give really fascinating perspectives to falsify
modified gravity theories and measuring the distance, the spin, or other physical parameters of the black holes. The geometries of the black holes allow to quantify the effect of the magnetic charge on deflection angle, and greybody factors our results lay to rest claims about strong clue of physical properties.


\begin{thebibliography}{83}
\bibitem{1} B.~P.~Abbott \textit{et al.} [LIGO Scientific and VIRGO],
``GW170104: Observation of a 50-Solar-Mass Binary Black Hole Coalescence at Redshift 0.2,''
Phys. Rev. Lett. \textbf{118}, no.22, 221101 (2017)
[erratum: Phys. Rev. Lett. \textbf{121}, no.12, 129901 (2018)].


\bibitem{4} K.~Akiyama \textit{et al.} [Event Horizon Telescope],
``First M87 Event Horizon Telescope Results. VI. The Shadow and Mass of the Central Black Hole,''
Astrophys. J. Lett. \textbf{875}, no.1, L6 (2019).



\bibitem{6} D.~Lovelock,
``The Einstein tensor and its generalizations,''
J. Math. Phys. \textbf{12}, 498-501 (1971).

\bibitem{7}  C.~Lanczos,
``A Remarkable property of the Riemann-Christoffel tensor in four dimensions,''
Annals Math. \textbf{39}, 842-850 (1938).

\bibitem{8} D.~Glavan and C.~Lin,
``Einstein-Gauss-Bonnet Gravity in Four-Dimensional Spacetime,''
Phys. Rev. Lett. \textbf{124}, no.8, 081301 (2020)

\bibitem{9}  P.~G.~S.~Fernandes,
``Charged black holes in AdS spaces in 4D Einstein Gauss-Bonnet gravity,''
Phys. Lett. B \textbf{805}, 135468 (2020)

\bibitem{10} A.~\"Ovg\"un,
``Black hole with confining electric potential in scalar-tensor description of regularized 4-dimensional Einstein-Gauss-Bonnet gravity,''
Phys. Lett. B \textbf{820}, 136517 (2021).

\bibitem{11}D.~V.~Singh, B.~K.~Singh and S.~Upadhyay,
``4$D$ AdS Einstein\textendash{}Gauss\textendash{}Bonnet black hole with Yang\textendash{}Mills field and its thermodynamics,''
Annals Phys. \textbf{434}, 168642 (2021).

\bibitem{12} 
S.~G.~Ghosh, D.~V.~Singh, R.~Kumar and S.~D.~Maharaj,
``Phase transition of AdS black holes in 4D EGB gravity coupled to nonlinear electrodynamics,''
Annals Phys. \textbf{424}, 168347 (2021).

\bibitem{Singh:2021iwv}
K.~N.~Singh, S.~K.~Maurya, A.~Dutta, F.~Rahaman and S.~Aktar,
``Quark stars in 4-dimensional Einstein\textendash{}Gauss\textendash{}Bonnet gravity,''
Eur. Phys. J. C \textbf{81}, no.10, 909 (2021)



\bibitem{14} R.~Kumar and S.~G.~Ghosh,
``Rotating black holes in $4D$ Einstein-Gauss-Bonnet gravity and its shadow,''
JCAP \textbf{07}, 053 (2020).



\bibitem{22} P.~G.~S.~Fernandes, P.~Carrilho, T.~Clifton and D.~J.~Mulryne,
``Derivation of Regularized Field Equations for the Einstein-Gauss-Bonnet Theory in Four Dimensions,''
Phys. Rev. D \textbf{102}, no.2, 024025 (2020)

\bibitem{23}B.~Eslam Panah, K.~Jafarzade and S.~H.~Hendi,
``Charged 4D Einstein-Gauss-Bonnet-AdS black holes: Shadow, energy emission, deflection angle and heat engine,''
Nucl. Phys. B \textbf{961}, 115269 (2020).




\bibitem{30} S.~S.~Li, S.~Mao, Y.~Zhao and Y.~Lu,
``Gravitational lensing of gravitational waves: A statistical perspective,''
Mon. Not. Roy. Astron. Soc. \textbf{476}, no.2, 2220-2229 (2018).

\bibitem{31} J. Soldner, Ueber die Ablenkung eines Lichtstrals von seiner ger adlinigen Bewegung, durch die Attraktion eines Weltkörpers, an welchem er nahe vorbei geht. Berliner Astronomisches Jahrbuch, 161-172.cc(1804).

\bibitem{32}M.~Bartelmann and P.~Schneider,
``Weak gravitational lensing,''
Phys. Rept. \textbf{340}, 291-472 (2001).

\bibitem{33} H.~C.~D.~L.~Junior, P.~V.~P.~Cunha, C.~A.~R.~Herdeiro and L.~C.~B.~Crispino,
``Shadows and lensing of black holes immersed in strong magnetic fields,''
Phys. Rev. D \textbf{104}, no.4, 044018 (2021)


\bibitem{35} M.~Sharif and S.~Iftikhar,
``Strong gravitational lensing in non-commutative wormholes,''
Astrophys. Space Sci. \textbf{357}, no.1, 85 (2015).



\bibitem{39} G.~W.~Gibbons and M.~C.~Werner,
``Applications of the Gauss-Bonnet theorem to gravitational lensing,''
Class. Quant. Grav. \textbf{25}, 235009 (2008).

\bibitem{40}M.~C.~Werner,
``Gravitational lensing in the Kerr-Randers optical geometry,''
Gen. Rel. Grav. \textbf{44}, 3047-3057 (2012). 


\bibitem{loboo}
K.~Jafarzade, M.~Kord Zangeneh and F.~S.~N.~Lobo,
``Shadow, deflection angle and quasinormal modes of Born-Infeld charged black holes,''
JCAP \textbf{04}, 008 (2021).



\bibitem{43} A.~\"Ovg\"un,
``Weak field deflection angle by regular black holes with cosmic strings using the Gauss-Bonnet theorem,''
Phys. Rev. D \textbf{99}, no.10, 104075 (2019)



\bibitem{45} K.~de Leon and I.~Vega,
``Weak gravitational deflection by two-power-law densities using the Gauss-Bonnet theorem,''
Phys. Rev. D \textbf{99}, no.12, 124007 (2019).




\bibitem{49}K.~Takizawa, T.~Ono and H.~Asada,
``Gravitational deflection angle of light: Definition by an observer and its application to an asymptotically nonflat spacetime,''
Phys. Rev. D \textbf{101}, no.10, 104032 (2020).


\bibitem{51} Z.~Li and A.~\"Ovg\"un,
``Finite-distance gravitational deflection of massive particles by a Kerr-like black hole in the bumblebee gravity model,''
Phys. Rev. D \textbf{101}, no.2, 024040 (2020).

\bibitem{52} Z.~Li, G.~Zhang and A.~\"Ovg\"un,
``Circular Orbit of a Particle and Weak Gravitational Lensing,''
Phys. Rev. D \textbf{101}, no.12, 124058 (2020).



\bibitem{56}K.~Jusufi and A.~\"Ovg\"un,
``Gravitational Lensing by Rotating Wormholes,''
Phys. Rev. D \textbf{97}, no.2, 024042 (2018).


\bibitem{59} A.~\"Ovg\"un,
``Deflection Angle of Photons through Dark Matter by Black Holes and Wormholes Using Gauss\textendash{}Bonnet Theorem,''
Universe \textbf{5}, no.5, 115 (2019).



\bibitem{Kumaran:2021rgj}
Y.~Kumaran and A.~\"Ovg\"un,
``Deriving Weak Deflection Angle by Black Holes or Wormholes using Gauss-Bonnet Theorem,''
Turk. J. Phys. \textbf{45}, 247-267 (2021)

\bibitem{Okyay:2021nnh}
M.~Okyay and A.~\"Ovg\"un,
``Nonlinear electrodynamics effects on the black hole shadow, deflection angle, quasinormal modes and greybody factors,''
JCAP \textbf{01}, no.01, 009 (2022).

\bibitem{Pantig:2021zqe}
R.~C.~Pantig, P.~K.~Yu, E.~T.~Rodulfo and A.~\"Ovg\"un,
``Shadow and weak deflection angle of extended uncertainty principle black hole surrounded with dark matter,''
Annals of Physics 436, 168722 (2022).

\bibitem{Javed:2020pyz}
W.~Javed, J.~Abbas, Y.~Kumaran and A.~\"Ovg\"un,
``Weak deflection angle by asymptotically flat black holes in Horndeski theory using Gauss-Bonnet theorem,''
Int. J. Geom. Meth. Mod. Phys. \textbf{18}, no.01, 2150003 (2021).

\bibitem{Ovgun:2020yuv}
A.~\"Ovg\"un,
``Weak Deflection Angle of Black-bounce Traversable Wormholes Using Gauss-Bonnet Theorem in the Dark Matter Medium,''
Turk. J. Phys. \textbf{44}, no.5, 465-471 (2020)



\bibitem{Takizawa:2020dja}
K.~Takizawa, T.~Ono and H.~Asada,
``Gravitational lens without asymptotic flatness: Its application to the Weyl gravity,''
Phys. Rev. D \textbf{102}, no.6, 064060 (2020).



\bibitem{Ishihara:2016vdc}
A.~Ishihara, Y.~Suzuki, T.~Ono, T.~Kitamura and H.~Asada,
``Gravitational bending angle of light for finite distance and the Gauss-Bonnet theorem,''
Phys. Rev. D \textbf{94}, no.8, 084015 (2016).


\bibitem{60} A.~\"Ovg\"un,
``Light deflection by Damour-Solodukhin wormholes and Gauss-Bonnet theorem,''
Phys. Rev. D \textbf{98}, no.4, 044033 (2018).

\bibitem{61}W.~Javed, R.~Babar and A.~\"Ovg\"un,
``The effect of the Brane-Dicke coupling parameter on weak gravitational lensing by wormholes and naked singularities,''
Phys. Rev. D \textbf{99}, no.8, 084012 (2019)

\bibitem{62} W.~Javed, R.~Babar and A.~\"Ovg\"un,
``Effect of the dilaton field and plasma medium on deflection angle by black holes in Einstein-Maxwell-dilaton-axion theory,''
Phys. Rev. D \textbf{100}, no.10, 104032 (2019).

\bibitem{63} W.~Javed, J.~Abbas and A.~\"Ovg\"un,
``Deflection angle of photon from magnetized black hole and effect of nonlinear electrodynamics,''
Eur. Phys. J. C \textbf{79}, no.8, 694 (2019).

\bibitem{64}W.~Javed, j.~Abbas and A.~\"Ovg\"un,
``Effect of the Hair on Deflection Angle by Asymptotically Flat Black Holes in Einstein-Maxwell-Dilaton Theory,''
Phys. Rev. D \textbf{100}, no.4, 044052 (2019).

\bibitem{65}W.~Javed, A.~Hamza and A.~\"Ovg\"un,
``Effect of nonlinear electrodynamics on the weak field deflection angle by a black hole,''
Phys. Rev. D \textbf{101}, no.10, 103521 (2020).


\bibitem{66} W.~Javed, M.~B.~Khadim, A.~\"Ovg\"un and J.~Abbas,
``Weak gravitational lensing by stringy black holes,''
Eur. Phys. J. Plus \textbf{135}, no.3, 314 (2020).

\bibitem{67} W.~Javed, M.~B.~Khadim and A.~\"Ovg\"un,
``Weak gravitational lensing by Bocharova\textendash{}Bronnikov\textendash{}Melnikov\textendash{}Bekenstein black holes using Gauss\textendash{}Bonnet theorem,''
Eur. Phys. J. Plus \textbf{135}, no.7, 595 (2020).



\bibitem{74} L.~Tannukij, P.~Wongjun and S.~G.~Ghosh,
``Black String in dRGT Massive Gravity,''
Eur. Phys. J. C \textbf{77}, no.12, 846 (2017).

\bibitem{75}R.~A.~Konoplya and A.~F.~Zinhailo,
``Grey-body factors and Hawking radiation of black holes in $4D$ Einstein-Gauss-Bonnet gravity,''
Phys. Lett. B \textbf{810}, 135793 (2020).

\bibitem{83}P.~Boonserm and M.~Visser,
``Bounding the greybody factors for Schwarzschild black holes,''
Phys. Rev. D \textbf{78}, 101502 (2008).




\bibitem{77} P.~Boonserm, T.~Ngampitipan and M.~Visser,
``Bounding the greybody factors for scalar field excitations on the Kerr-Newman spacetime,''
JHEP \textbf{03}, 113 (2014).



\bibitem{Ngampitipan:2013sf}
T.~Ngampitipan and P.~Boonserm,
``Bounding the greybody factors for the Reissner-Nordstr\"om black holes,''
J. Phys. Conf. Ser. \textbf{435}, 012027 (2013).

\bibitem{Mistry} R.~Mistry \textit{et al.}, ``Hawking radiation power equations for black holes'', Nuclear Phys. B \textbf{923}, 378–393 (2017).

\bibitem{79} S.~I.~Kruglov,
``New model of 4D Einstein--Gauss--Bonnet gravity coupled with nonlinear electrodynamics,''
Universe \textbf{7}, 249 (2021).

\bibitem{Qi}  F.~Q.~Ming, Li~Zhao and Y.~X.~ Liu, ``Weak deflection angle by electrically and magnetically charged black holes from nonlinear electrodynamics'',         	Phys. Rev. D \textbf{104}, 024033 (2021).

\bibitem{80} G.~Crisnejo and E.~Gallo,
``Weak lensing in a plasma medium and gravitational deflection of massive particles using the Gauss-Bonnet theorem. A unified treatment,''
Phys. Rev. D \textbf{97}, no.12, 124016 (2018).




\end{thebibliography}
\end{document}